\documentclass[english,10pt,pre,aps,showpacs,twocolumn,amsmath,amssymb,superscriptaddress,longbibliography,floatfix]{revtex4-1}
\usepackage[T1]{fontenc}
\usepackage[latin9]{inputenc}
\usepackage{verbatim}
\usepackage{refstyle}
\usepackage{amsmath}
\usepackage{amssymb}
\usepackage{graphicx}
\usepackage{bm}
\usepackage{times}
\usepackage{hyperref}
\hypersetup{pdftex,colorlinks=true,allcolors=blue}
\usepackage{hypcap}
\makeatletter

\AtBeginDocument{\providecommand\eqref[1]{\ref{eq:#1}}}
\AtBeginDocument{}
\AtBeginDocument{}
\RS@ifundefined{subsecref}
  {\newref{subsec}{name = \RSsectxt}}
  {}
\RS@ifundefined{thmref}
  {\def\RSthmtxt{theorem~}\newref{thm}{name = \RSthmtxt}}
  {}
\RS@ifundefined{lemref}
  {\def\RSlemtxt{lemma~}\newref{lem}{name = \RSlemtxt}}
  {}

\usepackage[color=green!40]{todonotes}

\usepackage{braket}

\newcommand{\tr}{\mathop{\mathrm{Tr}}}
\newcommand{\nod}[1]{:\! #1 \!:}
\makeatother

\usepackage{babel}
\begin{document}

\title{Non-Gibbs states on a Bose-Hubbard lattice}

\author{Alexander Yu.~Cherny}
\email{cherny@theor.jinr.ru}
\affiliation{Bogoliubov Laboratory of Theoretical Physics, Joint Institute for Nuclear
Research, 141980, Dubna, Moscow region, Russia}
\affiliation{Center for Theoretical Physics of Complex Systems, Institute for Basic Science (IBS), Daejeon 34051, Republic of Korea}

\author{Thomas Engl}
\affiliation{Center for Theoretical Physics of Complex Systems, Institute for Basic Science (IBS), Daejeon 34051, Republic of Korea}
\affiliation{
New Zealand Institute for Advanced Study, and
Centre for Theoretical Chemistry and Physics, Massey University, Private Bag 102904 North Shore,  Auckland 0745, New Zealand}

\author{Sergej Flach}
\affiliation{Center for Theoretical Physics of Complex Systems, Institute for Basic Science (IBS), Daejeon 34051, Republic of Korea}
\affiliation{
New Zealand Institute for Advanced Study, and
Centre for Theoretical Chemistry and Physics, Massey University, Private Bag 102904 North Shore,  Auckland 0745, New Zealand}

\begin{abstract}
We study the equilibrium properties of the repulsive quantum Bose-Hubbard model at high temperatures in arbitrary dimensions, with and without disorder. In its microcanonical setting the model conserves energy and particle number. The microcanonical dynamics is characterized by a pair of two densities: energy density $\varepsilon$ and particle number density  $n$. The macrocanonical Gibbs distribution also depends on two parameters: the inverse nonnegative temperature $\beta$ and the chemical potential $\mu$. We prove the existence of non-Gibbs states, that is, pairs $(\varepsilon,n)$ which cannot be mapped onto $(\beta,\mu)$. The separation line in the density control parameter space between Gibbs and non-Gibbs states $\varepsilon \sim n^2$ corresponds to infinite temperature $\beta=0$. The non-Gibbs phase cannot be cured into a Gibbs one within the standard Gibbs formalism using negative temperatures.
\end{abstract}
\maketitle

\section{Introduction}
\label{sec:intro}

Equipartition, ergodicity, and thermalization are essential properties, which an isolated many-body system has to have
in order to qualify for the applicability of fundamental laws of statistical mechanics.
Violation of the former, on the other side, could preserve coherence and might be of interest for, e.g., efficient information processing on classical and quantum levels. Surprising indications of transitions from ergodic to nonergodic dynamics have been reported in models of Josephson junction chains \cite{Pino:2016} and Bose-Einstein condensates of ultracold atoms on optical lattices \cite{PhysRevLett.120.184101} upon \emph{``heating''} the systems, i.e., upon {\sl increasing} the average unbounded energy density.

For a macroscopic system with the energy being the only relevant conserved quantity --- as in the case of the Josephson junction network \cite{Pino:2016} --- the canonical distribution function allows us to map any average energy density $\varepsilon$ into a positive inverse temperature $\beta$ of the canonical distribution \footnote{Note that this will hold independently of whether the microcanonical dynamics is ergodic (thermalizing) or nonergodic (nonthermalizing).}.

In the presence of a second conserved quantity (e.g., the particle number) the situation changes. The additional constraint unfolds the existence of a non-Gibbs statistics of the one-dimensional Gross-Pitaevskii (GP) lattice also known as the discrete nonlinear Schr\"odinger equation with Hamiltonian
\begin{equation}
H = \sum_l   -J\left( \psi_l^*\psi_{l+1} + \mathrm{c.c.} \right) +\frac{U}{2} |\psi_l|^4
\label{GPH}
\end{equation}
and the Hamiltonian equations of motion $ i \dot{\psi}_l = \partial H / \partial \psi_l^*$,
as obtained in Ref.~\cite{rasmussen00}.
The GP dynamics preserves the energy $H$ and the total norm $N = \sum_l |\psi_l|^2$. The microcanonical dynamics
at equilibrium --- if existent --- is defined by the two densities $\varepsilon=H/M$ and $n=N/M$, where $M$ is the number of lattice sites~\cite{note}.
In the macroscopic limit $M \rightarrow \infty$ it follows that the Gibbs grand-canonical formalism becomes only applicable to the microcanonical dynamics
for energy densities $\varepsilon \leqslant U n^2$, despite the fact that the microcanonical dynamics can address states with $\varepsilon > U n^2$. These states are therefore
called non-Gibbs states \cite{rasmussen00}. Subsequent studies addressed the question whether the microcanonical GP dynamics in the non-Gibbs phase is nonergodic
\cite{rumpf01,rumpf04,Rumpf2007,Rumpf2008,rumpf09,buonsante17}. Recent data \cite{PhysRevLett.120.184101} show that the dynamics stays
ergodic; however, relaxation times quickly grow deep in the non-Gibbs phase, turning the system into a dynamical glass, which becomes quickly nonergodic
for any practical purpose. The microscopic mechanism is related to the excitation of long-lived discrete breathers \cite{Flach:1998,flach08}. These discrete
breathers become precise single-action excitations of the system in the infinite density limit, where the GP system turns into an integrable set of uncoupled
anharmonic oscillators. The anomalously growing lifetime of these objects is the reason for the growth of relaxation times \cite{PhysRevLett.120.184101}.

In this work we consider the corresponding quantum model, the Bose-Hubbard model on lattices with arbitrary dimension $d$ (note that the particular choice of
lattice symmetry will not be of importance) \cite{fisher89}. For simplicity we only show here its direct one-dimensional nearest-neighbour counterpart of the GP lattice (\ref{GPH}):
\begin{equation}
\hat{H}=\sum_{l} -J\left( \hat{a}_{l}^{\dagger}\hat{a}_{l+1} + \mathrm{H.c.}\right) +\frac{U}{2}\hat{n}_{l}\left(\hat{n}_{l}-1\right).\label{eq:BH}
\end{equation}
Here,  $\hat{a}_{l}^{\dagger}$
and $\hat{a}_{l}$ are the bosonic creation and annihilation operators
on the $l$th lattice site with the commutation relations $[\hat{a}_{i},\hat{a}_{j}^{\dag}]=\delta_{ij}$, and $\hat{n}_{l}=\hat{a}_{l}^{\dagger}\hat{a}_{l}$
is the site occupation number.
The GP Hamiltonian (\ref{GPH}) is the classical limit of the Bose-Hubbard (BH) Hamiltonian (\ref{eq:BH}) for large occupation numbers $n_l\gg 1$, since quantum operators can be replaced by $c$-numbers: $\hat{a}_{i} \to \psi_{i}$, $\hat{a}_{i}^{\dag} \to \psi_{i}^{*}$.

We will show that the non-Gibbs phase exists in the full quantum BH lattice model as well
for energy densities $\varepsilon>Un^{2}$ in the $n$-$\varepsilon$
plane, where $n=N/M$ and $\varepsilon=E/M$ are the average occupation
number (filling factor) and energy per site, respectively, and $U \geqslant 0$ is the on-site
repulsion energy of the BH lattice Hamiltonian \cite{fisher89}.
We will also generalize
to disorder potentials. Our general proof of the existence of nonthermal non-Gibbs states
is relevant for a variety of experimental setups of Bose-Einstein condensates of ultracold atomic gases in optical lattice potentials, which are often used to study fundamental properties of matter \cite{RevModPhys.80.885}.

This paper is organized as follows: in Sec.~\ref{sec:mainres}, we obtain the line of infinite temperatures and, by means of the method of cumulant expansions, the thermodynamic susceptibilities for the Bose-Hubbard model. In the next section, we generalize these results to the Bose-Hubbard model with disorder (Sec.~\ref{sec:disorder}) and with arbitrary local interactions (Sec.~\ref{sec:extentions}). We consider the density of states of the system near the infinite temperature line and discuss the important role of boundedness of the single-particle spectrum. In the Conclusions, we review the obtained results and discuss possible applications and prospects.

\section{Main results}
\label{sec:mainres}

\subsection{The method}

The limiting line $\varepsilon=Un^{2}$, found in Ref.~\cite{rasmussen00},
corresponds to infinite temperatures of the canonical Gibbs ensemble.
However, the non-Gibbs region above the line in the $n$-$\varepsilon$
parameter space is experimentally (numerically) accessible, since
it is possible to initialize the lattice at any energy and total number
of particles \cite{rasmussen00,flach08}. For instance, putting all
$N$ bosons onto one site gives the energy per site $\varepsilon \sim N \sim M$,
which is infinitely large in the thermodynamic limit.
Dissipative boundaries were also proposed in order to drive the system into a non-Gibbs region \cite{PhysRevLett.97.060401,1367-2630-15-2-023032}.
Here we address the question whether the limiting line also exists in the \emph{quantum}
Bose-Hubbard model and which equation it obeys.

In order to answer this question, we calculate the average
energy in the canonical ensemble, that is, the internal energy as
a function of the temperature, total number of particles, and the
number of lattice sites, $E=\langle\hat{H}\rangle=E\left(T,N,M\right)=M\varepsilon\left(T,n\right)$.
The last equality is valid in the thermodynamic limit, because the
internal energy and total number of particles are extensive properties
of the system. The Gibbs canonical ensemble is, conceptually, the
most convenient one to study this problem, because we know the full range
of values of thermodynamic variables $T$ and $n$. For a given density
$n$ and for $0\leqslant T<\infty$, the line $\varepsilon\left(T,n\right)$
goes through all possible values of $\varepsilon$ in the $n$-$\varepsilon$
plane. The capacity
\begin{equation}
\left(\partial E/\partial T\right)_{N,M}=M\left(\partial\varepsilon/\partial T\right)_{n}\label{eq:cap}
\end{equation}
is positive, which implies that the energy reaches its maximum for
$T\to+\infty$, and, therefore, $\varepsilon(n)=\lim_{T\to+\infty}\varepsilon\left(T,n\right)$
is the upper bound for the Gibbs region in the $n$-$\varepsilon$
plane, provided the limit exists.

The grand canonical potential is defined as $\Omega=\Omega\left(T,\mu,M\right)=E-TS-\mu N$
with $S$ being the entropy, and $\mu=\left(\partial E/\partial N\right)_{T,M}=\left(\partial\varepsilon/\partial n\right)_{T}$
is the chemical potential. Below we use two independent variables
$\beta\equiv1/T$ and $\alpha\equiv-\beta\mu$ instead of $T$ and
$\mu$ to simplify calculations. For the grand canonical Gibbs ensemble,
the partition function $Z\left(\beta,\alpha,M\right)=\exp\left(-\beta\Omega\right)$
is given by
\begin{equation}
Z\left(\beta,\alpha,M\right)=\tr\exp\left(-\beta\hat{H}-\alpha\hat{N}\right).\label{eq:GCpf}
\end{equation}
Then the internal energy and total number of particles per site take the form
\begin{align}
\varepsilon\left(\beta,\alpha\right) & =-\frac{1}{M}\frac{\partial\ln Z\left(\beta,\alpha,M\right)}{\partial\beta},\label{eq:gE}\\
n\left(\beta,\alpha\right) & =-\frac{1}{M}\frac{\partial\ln Z\left(\beta,\alpha,M\right)}{\partial\alpha}.\label{eq:gN}
\end{align}
The grand canonical ensemble is equivalent to the canonical one only for nonzero
susceptibility $\left(\partial N/\partial\mu\right)_{\beta}$, since otherwise
the chemical potential is not a well-defined function of $n$. The susceptibility vanishes
at \emph{low} temperatures for the Bose Hubbard model
when the on-site repulsion $U/J$ is sufficiently large \cite{fisher89}.
As a consequence the system becomes a Mott insulator with zero compressibility.
However, at \emph{high} temperatures, the susceptibility is not zero,
as shown below.

Excluding the variable $\alpha$ from Eqs.~(\ref{eq:gE}) and (\ref{eq:gN})
yields $\varepsilon$ as a function of temperature and filling factor
and hence the particular case $\varepsilon=\varepsilon(n)$ at $\beta=0$, which we are interested in.

\subsection{Perturbation series at high temperatures}
\label{sec:pert}

For the GP lattice, the value $\alpha\left(\beta,n\right)=-\beta\mu\left(\beta,n\right)$ is shown to be \emph{finite and positive} in the limit $\beta\to 0$ ~\cite{rasmussen00}. We assume that this property is also valid for the BH model and then show the consistency
of the assumption. Thus, the fugacity
$z={e}^{\beta\mu}={e}^{-\alpha}$ (see, e.g., Ref.~\cite{book:huang87}) will tend
to a constant value less than one in the same limit.

In order to obtain a perturbation series at high temperatures for
the grand partition function, one can use the commutativity $\hat{H}$
and $\hat{N}=\sum_{i}\hat{n}_{i}$ and expand
$\exp\left(-\beta\hat{H}\right)$ in a power series in the inverse
temperature
\begin{align}
Z(\beta,\alpha,&M)  =\tr\left[\left(1-\beta\hat{H}+\beta^{2}\hat{H}^{2}/2+\cdots\right)\exp\left(-\alpha\hat{N}\right)\right]\nonumber \\
 & =Z_{0}\left(\alpha,M\right)\left(1-\beta\big\langle \hat{H}\big\rangle _{0}+\beta^{2}\big\langle \hat{H}^{2}\big\rangle _{0}/2+\cdots\right)
\end{align}
thus arriving at the expansion with the statistical moments $\mu_{m}\equiv\big\langle \hat{H}^{m}\big\rangle _{0}$.
Here we denote $Z_{0}\left(\alpha,M\right)=\tr\exp\left(-\alpha\hat{N}\right)$
and the brackets $\big\langle \cdots\big\rangle _{0}=\tr\left(\cdots\hat{\rho}_{0}\right)$
stand for the average over the density matrix in the zero-order approximation
\begin{align}
\hat{\rho}_{0}=\frac{\exp\left(-\alpha\sum_{i}\hat{n}_{i}\right)}{Z_{0}\left(\alpha,M\right)}. \label{eq:rho0}
\end{align}

The usage of the moment expansion is not convenient, since $\ln Z\left(\beta,\alpha,M\right)$
should be proportional to $M$ in the thermodynamic limit, while the
moment $\mu_{m}$ is proportional
to $M^{m}$.
In order to solve this problem, we reexpand $\ln\left(1-\beta\mu_{1}+\beta^{2}\mu_{2}/2!-\beta^{3}\mu_{3}/3!+\cdots\right)$
in powers of $\beta$ thus arriving at the \emph{cumulant expansion} with respect to $-\beta$ \cite{book:rushbrooke74}
\begin{equation}
\ln Z\left(\beta,\alpha,M\right)=
\lambda_{0}-\beta\frac{\lambda_{1}}{1!}+\beta^{2}\frac{\lambda_{2}}{2!}-\beta^{3}\frac{\lambda_{3}}{3!}+\cdots
\end{equation}
where each cumulant $\lambda_i$ depends on $\alpha$ and $M$,
and $\lambda_{0}=\ln Z_{0}\left(\alpha,M\right)$. Cumulants
can be obtained from the moments
\begin{equation}
\lambda_{1}=\mu_{1},\ \lambda_{2}=\mu_{2}-\mu_{1}^{2},\ \lambda_{3}=\mu_{3}-3\mu_{1}\mu_{2}+2\mu_{1}^{3},\label{eq:cumexpl}
\end{equation}
and so on. Since $\ln Z$ and all its derivatives
with respect to $\beta$ are proportional to $M$ in the thermodynamic
limit, the terms of order $M^{2}$ and higher cancel each other, and
each cumulant turns out to be proportional to $M$.

When calculating the energy per site at infinite temperatures,
it is sufficient to restrict ourselves to the first cumulant.
For obtaining the capacity (\ref{eq:cap}), we need the second order expansion
\begin{align}
\ln Z\left(\beta,\alpha,M\right)= & \ln Z_{0}\left(\alpha,M\right)-\beta\big\langle \hat{H}\big\rangle _{0}\nonumber \\
 & +\beta^{2}\left(\big\langle\hat{H}^{2}\big\rangle _{0}-\big\langle \hat{H}\big\rangle _{0}^{2}\right)/2+\cdots.
\label{eq:pf2dcum}
\end{align}

\subsection{The high-temperature expansion for the Bose-Hubbard model \label{sec:htexp}}

We calculate the partition function in the zero-order approximation
in the Fock basis $\Ket{\ldots,n_{i},\ldots}$ for the operators $\hat{a}_{i}$
and $\hat{a}_{i}^{\dag}$, where the density matrix is diagonal,
\begin{equation}
Z_{0}\left(\alpha,M\right)=\left[\sum_{n_{i}=0}^{\infty}\exp\left(-\alpha n_{i}\right)\right]^{M}=\frac{1}{\left(1-{e}^{-\alpha}\right)^{M}}.\label{eq:Z0}
\end{equation}

Since the density matrix (\ref{eq:rho0}) is the exponential of a quadratic form of the bosonic operators $\hat{a}_{i}^{\dag}$
and $\hat{a}_{i}$, the Wick-Bloch-De Dominicis  theorem \cite{bloch58} can be applied for calculating the average of a product of an arbitrary number of bosonic operators. According to the theorem, this average is given by the complete sum of all possible products of elementary \emph{binary} averages (contractions)
\begin{align}
\langle\hat{a}_{i}^{\dagger}\hat{a}_{j}\rangle _{0}&=\delta_{ij}n_0. \label{eq:aiajav}
\end{align}
Here $n_0$ is the mean occupation number per site (lattice filling factor) at $\beta=0$,
\begin{align}
n_{0}=\langle\hat{a}_{i}^{\dagger}\hat{a}_{i}\rangle_{0}=\left(1-{e}^{-\alpha}\right)\sum_{k=0}^{\infty}k\,{e}^{-\alpha k}
=\frac{1}{{e}^{\alpha}-1},\label{n0}
\end{align}
and the ``nondiagonal'' values are zero, since the average $\Braket{\ldots,n_{k},\ldots|\hat{a}_{j}^{\dagger}\hat{a}_{i}|\ldots,n_{k},\ldots}$ vanishes for $i\neq{}j$.

The mean value of the energy (\ref{eq:BH}) at $\beta=0$ is obtained as follows. The hopping terms in $\big\langle \hat{H}\big\rangle _{0}$ vanish, since they include only nondiagonal binary terms.
We are left only with the interaction energy terms, for which in accordance with the Wick--Bloch--De Dominicis  theorem, $\langle\hat{a}_{i}^{\dagger 2}\hat{a}_{i}^2\rangle_0=2\langle\hat{a}_{i}^{\dagger}\hat{a}_{i}\rangle_0^2=2n_0^2$. Thus
\begin{equation}
\big\langle \hat{H}\big\rangle_{0}
={MU}n_{0}^2.
\label{eq:Eav0}
\end{equation}

We substitute Eqs.~(\ref{eq:Z0}) and (\ref{eq:Eav0}) into Eq.~(\ref{eq:pf2dcum}) and
use Eqs.~(\ref{eq:gE}) and (\ref{eq:gN}) to derive
\begin{align}
\varepsilon & ={U}{n_{0}^2}+{O}\left(\beta\right),\label{eq:gE1}\\
n& =n_{0}[1-2\beta U n_{0}(n_{0}+1)]+{O}\left(\beta^{2}\right).\label{eq:gN1}
\end{align}
Here the symbol ${O}\left(x\right)$ denotes terms of
order $x$ and higher.

This system of equations allows us to obtain the upper border for the
Gibbs region in the $n$-$\varepsilon$ plane. For infinite temperatures, we have $n=n_{0}$ and
\begin{equation}
\varepsilon=Un^{2},\label{eq:enborder}
\end{equation}
which is the main result of this work.
It coincides with the Gibbs--non-Gibbs border for the discrete nonlinear Schr\"odinger
equation \cite{rasmussen00}.

In the limit of infinite temperatures, the average energy per particle (18) is finite at fixed particle density. In the same limit, the density matrix (8) is independent of the energy, which makes all energies equiprobable at a fixed particle density. On the other hand, the upper bound of the spectrum of the Bose-Hubbard Hamiltonian grows as $M^2$, which seemingly leads to a divergence of the average energy in the thermodynamic limit.
The resolution of this apparent paradox is that \emph{the density of states} decreases rapidly at high energies.

Equations (\ref{n0}) and (\ref{eq:gN1}) yield the parameter $\alpha$ up to the first-order term proportional to $\beta$:
\begin{equation}
\mu=-\frac{1}{\beta}\ln\left(1+1/n\right)+2Un+{O}\left(\beta\right) \label{eq:mu1}
\end{equation}
which leads to
\begin{equation}
\left(\frac{\partial\mu}{\partial n}\right)_{\beta}=\frac{1}{\beta n\left(n+1\right)}+2U+{O}\left(\beta\right).\label{eq:nbeta}
\end{equation}
The quantity (\ref{eq:nbeta}) is directly related to the fluctuations of the total
number of particles in the grand canonical ensemble, which remain
\emph{finite} in the limit of infinite temperatures,
\begin{align}
\frac{\big\langle \hat{N}^{2}\big\rangle -\big\langle \hat{N}\big\rangle^{2}}{\big\langle \hat{N}\big\rangle }
=& \frac{1}{\beta n}\left(\frac{\partial n}{\partial \mu}\right)_{\beta}=(n+1) \nonumber\\
&\times\left[1-2\beta n(n+1)U+O(\beta^2)\right].
\label{eq:nfluct}
\end{align}

In order to obtain the capacity (\ref{eq:cap}), we have to
consider the second-order terms in Eq.~(\ref{eq:pf2dcum}). In the same
manner as for Eq.~(\ref{eq:Eav0}), we derive
\begin{equation}
\big\langle \hat{H}^{2}\big\rangle _{0}-\big\langle \hat{H}\big\rangle_{0}^{2}=
Mn_{0}\left(n_{0}+1\right)\left[2J^{2}d+U^{2}n_{0}\left(5n_{0}+1\right)\right],\label{eq:efluct}
\end{equation}
where $d$ is the dimension of the lattice. We rewrite the capacity
per site in the more convenient form
\begin{equation}
\left(\frac{\partial\varepsilon}{\partial T}\right)_{n}=-\beta^{2}\left(\frac{\partial\varepsilon}{\partial\beta}\right)_{n}=-\frac{\beta^{2}}{\left(\frac{\partial n}{\partial\alpha}\right)_{\beta}}\frac{\partial\left(\varepsilon,n\right)}{\partial\left(\beta,\alpha\right)},\label{eq:cap1}
\end{equation}
 where we use the standard thermodynamic notation for the Jacobian determinant. By means of Eqs.~(\ref{eq:pf2dcum})--(\ref{eq:Eav0})
and (\ref{eq:efluct}), we arrive at the capacity per site in
the lowest order in $\beta$
\begin{equation}
\left(\frac{\partial\varepsilon}{\partial T}\right)_{n}=\beta^{2}n\left(n+1\right)\left[2J^{2}d+U^{2}n\left(n+1\right)\right]+{O}\left(\beta^{3}\right),\label{eq:cap_fin}
\end{equation}
which is positive for any nonzero $\beta$ and turns zero at infinite temperature.

The positiveness of the expressions (\ref{eq:nfluct}) and (\ref{eq:cap_fin}) implies that the system is stable at high temperatures.

The classical limit of the BH model, that is, the classical GP lattice can be evaluated in a similar manner for arbitrary lattice dimension by replacing sums by corresponding integrals. However, it is simpler to find the thermodynamic relations by considering the classical limit $n\gg 1$ in the obtained quantum system equations. For instance, Eq.~(\ref{n0}) becomes in this limit $n_0= 1/\alpha$, and therefore, $n= 1/\alpha$ at infinite temperatures in the classical case (see the discussion in Sec.~\ref{sec:GPline} below). Similarly we can replace $n(n+1)$ by $n^2$ in the equations for the susceptibility (\ref{eq:nbeta}) and capacity (\ref{eq:cap_fin}), in order to find their classical limits. We note that the thermodynamic relations of the quantum and classical cases differ significantly for small particle densities.

Let us emphasize that the classical limit is \emph{not} realized by a quantum system at infinite temperatures. Even on the infinite temperature line, the relation between the density of particles and fugacity ($z={e}^{-\alpha}$) varies: in the quantum model $n=z/(1-z)$, while in the classical one, $n=-1/\ln z$. This is because the energy remains \emph{finite} in the limit of infinite temperatures at a given particle density. Note that the classical case can be obtained from the quantum one in the limit $n\gg1$, but an inverse scheme does not exist.

\section{Generalizations}

\subsection{Adding disorder\label{sec:disorder}}

Let us consider the Bose-Hubbard model with \emph{disorder} by adding
to the Hamiltonian (\ref{eq:BH}) a disorder potential $\hat{H}_{\mathrm{dis}}=\sum_{i}\epsilon_{i}\hat{a}_{i}^{\dagger}\hat{a}_{i}$
with random on-site energies $\epsilon_{i}$, obeying some probability
density distribution (see, e.g., Ref.~\cite{flach16}). Their average
value $\overline{\epsilon}=\lim_{M\to\infty}M^{-1}\sum_{i}\epsilon_{i}$
is assumed to be zero, while the variance $\sigma_{\epsilon}$ is
finite.

The disorder potential does not change the upper bound for the Gibbs region
(\ref{eq:enborder}). Its contribution to the energy per site
is equal to zero in the limit of infinite temperatures: $\big\langle \hat{H}_{\mathrm{dis}}\big\rangle _{0}=n_{0}M^{-1}\sum_{i}\epsilon_{i}=0$.
However, the presence of disorder influences the second-order term
in the high temperature expansion (\ref{eq:pf2dcum}) into
\begin{align}
\big\langle \hat{H}^{2}\big\rangle_{0} -\big\langle \hat{H}\big\rangle_{0}^{2}= & Mn_{0}\left(n_{0}+1\right)\nonumber \\
 & \times\left[\sigma_{\epsilon}+2J^{2}d+U^{2}n_{0}\left(5n_{0}+1\right)\right],\label{eq:efluct1}
\end{align}
and, hence, the capacity per site
\begin{align}
\left(\frac{\partial\varepsilon}{\partial T}\right)_{n}\!\!=\beta^{2}f(n)+{O}\left(\beta^{3}\right),\label{eq:cap_fin1}
\end{align}
where we use the notation
\begin{align}
  f(n)=n\left(n+1\right)\left[\sigma_{\epsilon}+2J^{2}d+U^{2}n\left(n+1\right)\right].\label{eq:fn}
\end{align}
The border of the Gibbs region and the capacity
at high temperatures are independent of the correlations in disorder
$\big\langle \epsilon_{i}\epsilon_{i+m}\big\rangle =\lim_{M\to\infty}M^{-1}\sum_{i}\epsilon_{i}\epsilon_{i+m}$
for $m\neq0$.
Therefore uncorrelated disorder with finite variance will result in the same border between the Gibbs and the non-Gibbs regime given
by \eqref{enborder}. This result holds independent of the dimensionality of the lattice, and generalizes the previously obtained
consideration of a one-dimensional disordered GP lattice \cite{flach16}.

\subsection{The density of states in the vicinity of the Gibbs - non-Gibbs separation line}

Let us calculate the density of states in the vicinity of the border separating the Gibbs and non-Gibbs states.
Instead of following the route by
calculating the entropy near $\beta=0$ by means of the grand canonical potential: $S(T,\mu,M)=-(\partial \Omega/\partial T)_{\mu,M}$,
we use Eq.~(\ref{eq:cap_fin1}) for the capacity, since entropy and capacity per site are related by the equation $(\partial \varepsilon/\partial T)_n= T(\partial s/\partial T)_n$. Then the expression (\ref{eq:cap_fin1}) leads to $(\partial s/\partial \beta)_n=-\beta f(n)$ and after integration over $\beta$ to
\begin{align}
s=s_0(n)-\frac{\beta^2}{2}f(n) + {O}\left(\beta^{3}\right)\label{eq:sbn}
\end{align}
with $s_0(n)$ being the entropy per site at infinite temperature
\begin{align}
  s_0(n)=(n+1)\ln(n+1)-n\ln n.\label{eq:s0}
\end{align}
The last equation can be obtained directly from the density matrix (\ref{eq:rho0}) using the well-known relation for the entropy, $S_0=-\tr \hat{\rho}_{0} \ln \hat{\rho}_{0}$.

In the same manner, we derive from Eqs.~(\ref{eq:enborder}) and (\ref{eq:cap_fin1})
\begin{align}
\varepsilon=Un^2-\beta f(n) + {O}\left(\beta^{2}\right). \label{eq:epsbn}
\end{align}

Excluding the inverse temperature from Eqs.~(\ref{eq:sbn}) and (\ref{eq:epsbn}) yields
\begin{align}
s=(n+1)\ln(n+1)-n\ln n-\frac{\big(\varepsilon-Un^2\big)^2}{2f(n)},\label{eq:sen}
\end{align}
where $f(n)$ is given by Eq.~(\ref{eq:fn}). By using the relation $W={e}^S={e}^{sM}$, we arrive at the  density of states in the microcanonical ensemble near the Gibbs - non-Gibbs border
\begin{align}
W(E,N,M)=\frac{(M+N)^{M+N}}{N^N M^M}
         \exp\left[-\frac{M\big(\varepsilon-Un^2\big)^2}{2f(n)}\right],\label{eq:WENM}
\end{align}
where $n=N/M$ and $\varepsilon=E/M$. This is the main contribution to the asymptotics of the density of states in the thermodynamic limit $N/M=\mathrm{const}$, $M\to\infty$. The density of states reaches its maximum precisely at the border $\varepsilon=Un^2$.

The density of states for the discrete GP Hamiltonian (\ref{GPH})
\begin{align}
W_\mathrm{cl}(E,N,M)=\left(\frac{N{e}}{M}\right)^M
         \exp\left[-\frac{M\big(\varepsilon-Un^2\big)^2}{2f_\mathrm{cl}(n)}\right],\label{eq:WENMcl}
\end{align}
can be inferred from Eq.~(\ref{eq:WENM}) in the classical limit $n\gg 1$, where $f_\mathrm{cl}(n)=n^2\left(\sigma_{\epsilon}+2J^{2}d+U^{2}n^2\right)$. In the particular case $U=0$ (the ideal Bose gas), the parabolic dependence of the entropy on the energy in the microcanonical ensemble was obtained up to a coefficient in Ref.~\cite{buonsante17} by another method.

Note that for both quantum and classical cases, the entropy is monotonously increasing with $n$ at fixed $\varepsilon$, while it is nonmonotonous for increasing $\varepsilon$ at fixed $n$. This result is analogous to the one obtained in Ref.~\cite{rumpf04} for the one-dimensional classical GP lattice.

\subsection{Bounded single particle spectra and non-Gibbs states}

The Gibbs - non-Gibbs separation line (\ref{eq:enborder}) exists due to the single-particle dispersion being \emph{bounded},
e.g., as $-2J\cos q$ for the one-band Bose-Hubbard
model, \eqref{BH}.
Note that generalizations to a finite number of bands generated by more complex lattice structures are straightforward.
At variance, an infinite number of bands or the single-particle
dispersion, or simply the case of free particle $q^{2}/2m$,  leads to unbounded spectra.
As a consequence non-Gibbs states are renormalized to infinite values
in the $n$-$\varepsilon$ plane and vanish from any consideration. Indeed, in the particular case $U=0$,
we have in general an ideal Bose gas with the dispersion $J_{\lambda}(q)$
with $\lambda$ being the band number. Then the total number of particles
is given by the standard expression
\begin{equation}
N=\sum_{q,\lambda}1/\left[\exp\left(\beta J_{\lambda}(q)+\alpha\right)-1\right],\label{eq:multband}
\end{equation}
where the quasimomentum $q$ runs over $M$ uniformly distributed
points in the Brillouin zone. In the
case of a one-band structure, only one band, say, $\lambda=1$ contributes
to the sum in \eqref{multband}. Then the limit $\beta\to0$ yields
$N=M/\left({e}^{\alpha}-1\right)$ which allows for a solution of $\alpha$ as a function of $n=N/M$. However,
for an infinite number of bands,  $\sum_{\lambda=1}^{\infty}1/\left({e}^{\alpha}-1\right)$ diverges. Therefore the assumption about
the finiteness of $\alpha=-\beta\mu$ when $\beta\to0$ is not applicable
anymore.

\subsection{Further extensions}
\label{sec:extentions}

Non-Gibbs states can be obtained
as well for the following generalized Bose-Hubbard models
\begin{equation}
\hat{H}=\sum_{i\neq j}J\left(i-j\right)\hat{a}_{i}^{\dagger}\hat{a}_{j}+\sum_{i}u\left(\hat{n}_{i}\right).
\label{eq:BHgen}
\end{equation}
The sum in the hopping term includes all sites, and the single particle dispersion $J\left(q\right)=\sum_{m}J\left(m\right){e}^{iqm}$
is assumed to be bounded. The standard Bose-Hubbard Hamiltonian (\ref{eq:BH}) is a particular case of Eq.~(\ref{eq:BHgen}) with $J(m)=-J(\delta_{m,1}+\delta_{m,-1})$ and $u(x)=Ux(x-1)/2$.

If the condition
\begin{align}\label{eq:ancond}
\lim_{x\rightarrow +\infty} \frac{u\left(x\right)}{x} \rightarrow +\infty
\end{align}
is satisfied, then a non-Gibbs phase with anomalous scaling properties of the energy will emerge.
This phase cannot be described by a grand partition function with negative temperature due to the divergence of the former.
If on the other side Eq.(\ref{eq:ancond}) is not satisfied, i.e., $\lim_{x \rightarrow \infty} u(x)/x \rightarrow C$  with $ 0 \leqslant C < \infty$, then
the line $\varepsilon\left(n,\beta=0\right)$ becomes the border between
Gibbs states with positive and negative temperatures (see also Refs.~\cite{samuelsen13,buonsante17}).

\subsubsection{The Gibbs--non-Gibbs transition line for the generalized Bose-Hubbard model}

By complete analogy with Sec.~\ref{sec:htexp}, one can show that at infinite temperatures the hopping terms do not contribute to the average energy. Its value is determined by the interaction terms and calculated with the density matrix (\ref{eq:rho0})
\begin{equation}
\varepsilon=(1-{e}^{-\alpha})\sum_{m=0}^{+\infty}u(m){e}^{-\alpha m}.
\label{eq:gBH}
\end{equation}
The density of particles is still given by Eq.~(\ref{n0}). Substituting the exponential ${e}^{\alpha}=1+1/n$ into Eq.~(\ref{eq:gBH}), we arrive at the curve of the energy density in the $\varepsilon$-$n$ plane
\begin{equation}
\varepsilon(n)=\frac{1}{n+1}\sum_{m=0}^{+\infty}u(m)\bigg(\frac{n}{n+1}\bigg)^{m},\label{eq:bGnG}
\end{equation}
above which non-Gibbs states for the Hamiltonian (\ref{eq:BHgen}) emerge.
\begin{figure}
\includegraphics[width=.95\columnwidth]{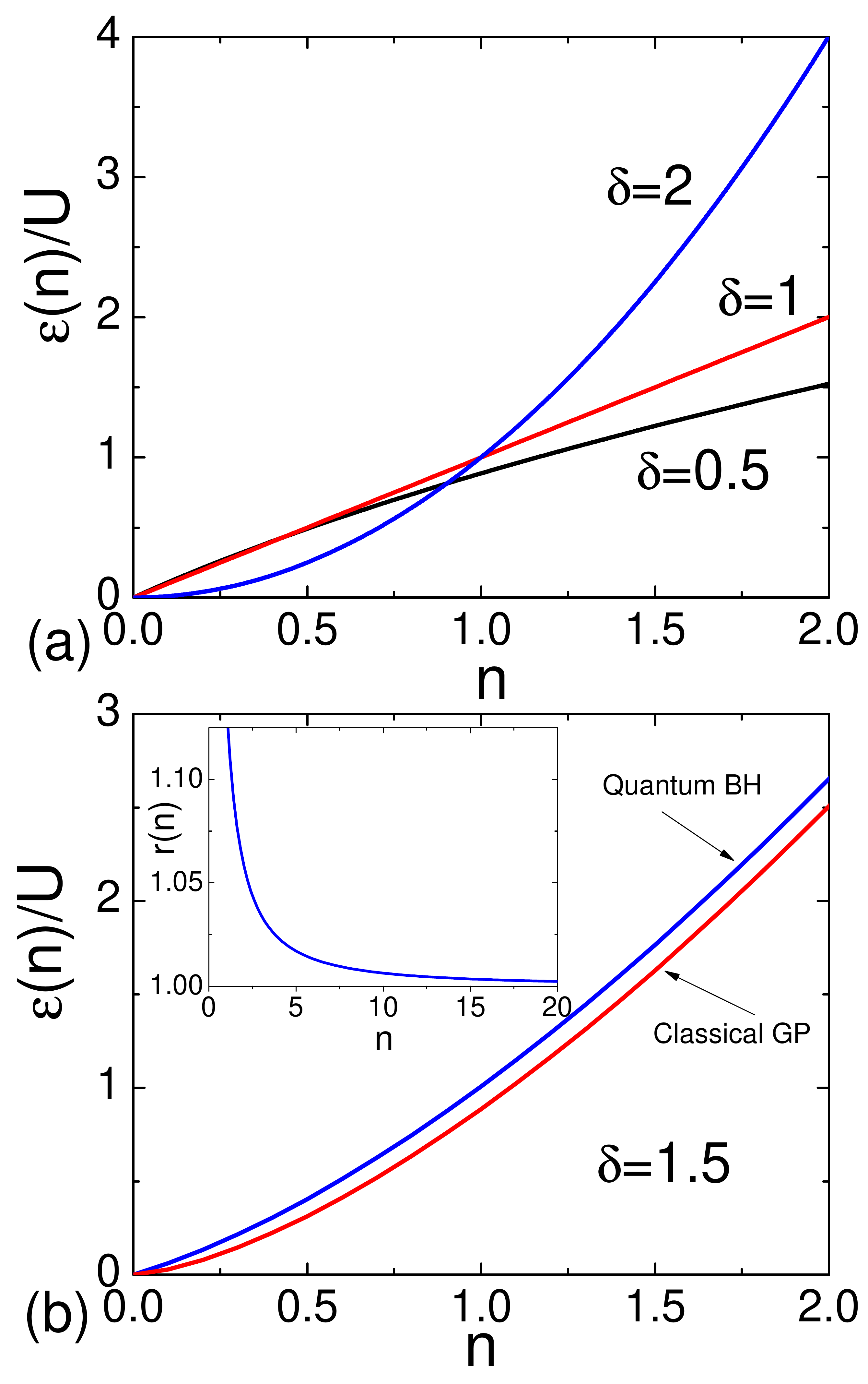}
\caption{\label{fig1} (Color online) (a) The infinite temperature line (\ref{bGnGpower}) in the $\varepsilon$-$n$ plane for the generalized Bose-Hubbard model (\ref{eq:BHgen}) with the local potential of the power-law type (\ref{eq:intpower}) for different values of the exponent $\delta$. Here $\varepsilon$ and $n$ are the energy and particle densities, respectively. Non-Gibbs states are located above the line if $\delta>1$. At variance, for $\delta\leqslant 1$ the line separates states with positive and negative temperatures.
Note that the case $\delta < 1$ corresponds to a class of Bose systems with saturable interactions satisfying $u(x)\to\infty$ and $\partial u(x)/\partial x\to0$ for $x\to\infty$.
(b) The infinite temperature lines $\varepsilon_{q,c}(n)$ for the quantum (\ref{eq:intpower}) and  classical (\ref{eq:uclass}) Bose-Hubbard and Gross-Pitaevskii models, respectively, as follows from Eqs.~(\ref{bGnGpower}) and  (\ref{bGnGpowerInt}). The interaction potential is chosen with noninteger parameter $\delta=3/2$.
Both lines differ at small densities, while their asymptotics for large $n$ coincides.
Inset: The ratio $r(n)=\varepsilon_q(n)/\varepsilon_c(n)$ of the quantum versus classical critical energy density as a function of particle density.  }
\end{figure}

As an example, we calculate the transition line for the local interaction energy
\begin{align}\label{eq:intpower}
u(x)=\frac{U}{\delta}\bigg[\frac{\Gamma(x+1)}{\Gamma(x+1-\delta)}-\frac{1}{\Gamma(1-\delta)}\bigg]
\end{align}
with $\Gamma(x)$ being Euler's gamma function. This form of the potential generalizes the power-law interaction $U\hat{a}^{\dagger m}\hat{a}^{m}/m$ on each site with a positive integer exponent $m$, for which $u(x)=Ux(x-1)\cdots (x-m+1)=\Gamma(x+1)/\Gamma(x+1-m)/m$. The last term in the right-hand side brackets in Eq.~(\ref{eq:intpower}) is needed to enforce the condition $u(0)=0$, which the potential should satisfy for any positive $\delta$. The long-range asymptotics of that potential is given by $u(x)=Ux^{\delta}/\delta[1+O(1/x)]$. The line of infinite temperatures is calculated by means of Eq.~(\ref{eq:bGnG}), which yields
\begin{align}
\varepsilon(n)=\frac{U}{\delta\Gamma(1-\delta)}\bigg[\frac{{}_{2}F_{1}(1,1;1-\delta;\frac{n}{n+1})}{n+1}-1\bigg],
\label{bGnGpower}
\end{align}
where ${}_{2}F_{1}(a,b;c;z)$ is the hypergeometric function~\cite{abr_steg64}. As discussed above, this line separates the Gibbs and non-Gibbs phases when $\delta>1$.
For positive integer exponent $\delta=1,2,\cdots$, Eq.~(\ref{bGnGpower}) is simplified
\begin{equation}
\varepsilon(n)=U\Gamma(\delta)n^{\delta}.
\label{bGnGpowerInt}
\end{equation}
This equation also gives the main term of the long-range asymptotics of Eq.~(\ref{bGnGpower}) for arbitrary positive $\delta$. For the standard Bose-Hubbard model with two-body interactions, we have $\delta=2$ and arrive at Eq.~(\ref{eq:enborder}) (see blue line in Fig.~\ref{fig1}). For $\delta=1$ the impact of the potential is reduced to a renormalization of the chemical potential of a noninteracting ideal Bose gas (see red line in Fig.~\ref{fig1}). As a result, this case separates
the appearance of non-Gibbs phases for $\delta > 1$ from cases with $\delta < 1$, where the infinite temperature line separates Gibbs states with positive and negative temperatures (see black line in Fig.~\ref{fig1}).

It is straightforward to extend the hopping network between
sites. For instance, for binary interactions a one-band Hamiltonian
takes the form
\begin{equation}
\hat{H}=\sum_{i\neq j}J\left(i-j\right)\hat{a}_{i}^{\dagger}\hat{a}_{j}+\frac{1}{2}\sum_{i,j}U\left(i-j\right)\hat{a}_{i}^{\dagger}\hat{a}_{j}^{\dagger}\hat{a}_{j}\hat{a}_{i}.\label{eq:BHgen1}
\end{equation}

Using the same method as in Sec.~\ref{sec:mainres}, we arrive at non-Gibbs states above the line $\varepsilon=U_{0}n^{2}$
with $U_{0}\equiv U(0)+\frac{1}{2}\sum_{j\not=0}U\left(j\right)$, which is assumed to be
finite and positive.

Similar to Sec.~\ref{sec:disorder}, we can show
that the disorder does not modify the border line between Gibbs
and non-Gibbs states for the both models, Eqs.~(\ref{eq:BHgen})
and (\ref{eq:BHgen1}).

\subsubsection{The Gibbs--non-Gibbs transition line for the discrete nonlinear Schr\"odinger equation}
\label{sec:GPline}

We also consider
the generalizations of the GP lattice (\ref{GPH}), whose Hamiltonian is given by Eq.~(\ref{eq:BHgen}) with the replacement $\hat{a}_{i} \to \psi_{i}$, $\hat{a}_{i}^{\dag} \to \psi_{i}^{*}$. One can calculate the Gibbs--non-Gibbs transition line in the same way as in Sec.~\ref{sec:mainres}. However, the easiest way is to obtain it in the limit of large occupation numbers, when the quantum model approaches the classical one. Then the density of particles in Eq.~(\ref{n0}) is $n=1/\alpha$ and the sum in Eq.~(\ref{eq:gBH}) tends to the integral $\int_{0}^{\infty} \mathrm{d} x u(x)\exp(-\alpha x)/\alpha$. We finally obtain
\begin{equation}
\varepsilon(n)=\frac{1}{n}\int_{0}^{\infty}u\left(x\right){e}^{-x/n}\mathrm{d} x\label{eq:bGnGcl}.
\end{equation}
We observe that the line of infinite temperatures $\varepsilon(n)$ is given by the  Laplace transform $F(s)$ of the interaction potential $u(x)$ through $\varepsilon(n) = s F(s) |_{s=1/n}$.
Similar to the quantum case, non-Gibbs states for the Hamiltonian (\ref{eq:BHgen}) emerge for energy densities $\varepsilon > \varepsilon(n)$ if the condition (\ref{eq:ancond}) is satisfied.

The classical analog of the potential (\ref{eq:intpower}) in the Bose-Hubbard model is
\begin{align}\label{eq:uclass}
u(x)=U x^{\delta}/\delta.
\end{align}
Using Eq.~(\ref{eq:bGnGcl}), we find the Gibbs--non-Gibbs separation line given by Eq.~(\ref{bGnGpowerInt}). Thus, for an integer exponent $\delta$, the
infinite temperature line for the quantum and classical models is identical, but for noninteger $\delta$ it is not, see Fig.~\ref{fig1}b.

One can see that for the interaction $\sum_{i}\nod{u\left(\hat{n}_{i}\right)}$ with $u(x)$ being an entire function of $x$, the line $\beta=0$ matches Eq.~(\ref{eq:bGnGcl}). Here the symbols $\nod{\cdots}$ stand for the normal ordering of the Bose operators $\hat{a}_{i}$ and $\hat{a}_{i}^{\dagger}$. In order to prove this equation, it is sufficient to expand an arbitrary function $u\left(x\right)$ in a Taylor series and show the equation validity for each term of the series. Thus, we consider the particular case $u\left(x\right)=x^{m}$ for any nonnegative integer $m$, for which $\nod{\hat{n}^{m}}=\hat{a}^{\dagger m}\hat{a}^{m}$ (here we omit the lattice site index $i$ for simplicity). By applying the methods of Sec.~\ref{sec:mainres} and using the Wick-Bloch-De Dominicis  theorem, we arrive at $\varepsilon=\big\langle \hat{a}^{\dagger m}\hat{a}^{m}\big\rangle _{0}=m!n_{0}^{m}=m!n^{m}$. The last equality is valid in zeroth order in $\beta$. This coincides with the relation $\varepsilon=m!n^m$ given by Eq.~(\ref{eq:bGnGcl}).

Samuelsen \emph{et al.} \cite{samuelsen13} considered a one-dimensional chain with the saturable nonlinear potential $u\left(x\right)=\nu\ln\left(1+x\right)$ (with negative $\nu < 0$), and obtained the line
of infinite temperatures using a transfer integral approach:
\begin{equation}
\varepsilon=\nu\exp\left(1/n\right)E_{1}\left(1/n\right),\label{eq:pnsat}
\end{equation}
 where $E_{1}\left(z\right)=\int_{z}^{\infty}\mathrm{d} t\,{e}^{-t}/t$
is the exponential integral \cite{abr_steg64}. We note that our method is generating and therefore confirming that
result using the above relation (\ref{eq:bGnGcl}). We also note that our approach allows one to generalize the result (\ref{eq:pnsat}) to
arbitrary lattice dimensions.

\subsection{The absence of the Gibbs--non-Gibbs transition line in a Josephson junction array model}
\label{sec:PR}

Let us discuss why the Gibbs--non-Gibbs transition line is absent in a Josephson junction array
\begin{equation}\label{eq:PRH}
\hat{H}=\sum_{i}\frac{E_{\mathrm{C}}}{2}\hat{q}_{i}^2-E_{\mathrm{J}}\cos(\hat{\varphi}_{i}-\hat{\varphi}_{i+1}).
\end{equation}
This is a simple model describing the Josephson junction network of weakly coupled superconducting islands with the Josephson $E_{\mathrm{J}}$ and charging $E_{\mathrm{C}}$ energies (see, e.g., Ref.~\cite{Pino:2016}). The operators of phase $\hat{\varphi}_{i}$ and charge $\hat{q}_{i}$ of the superconducting islands obey the commutation relations $[\hat{q}_{i},{e}^{i\hat{\varphi}_{j}}] = \delta_{ij}{e}^{i\hat{\varphi}_{j}}$. They can be considered as the $z$-component of momentum operators $\hat{q}_{i} \rightarrow \hat{l}_{zi}=-i\frac{\partial}{\partial \varphi_{i}} $ and $\hat{\varphi}_{i} \rightarrow \varphi_{i}$ on the ring of unit radius.

The Hamiltonian (\ref{eq:PRH}) is derived from the Bose-Hubbard model (\ref{eq:BH}) in the limit of infinite densities \cite{Bruder05}. In the polar representation, the bosonic operators follow as
\begin{align}
\hat{a}_{i}={e}^{-i\hat{\varphi}_{i}}\sqrt{n+\delta\hat{n}_{i}}, \quad \hat{a}^\dag_{i}=\sqrt{n+\delta\hat{n}_{i}}{e}^{i\hat{\varphi}_{i}} \label{eq:aporlar}
\end{align}
with $\delta\hat{n}_{i}$ being the operator of deviation from the average occupation number per site. Since the density is assumed to be large, the operator $\delta\hat{n}_{i}$ takes now all integer values. The operators $\delta\hat{n}_{i}$ and $\hat{\varphi}_{i}$ obeys the same commutation relations as $\hat{q}_{i}$ and $\hat{\varphi}_{i}$.

Substituting the operators (\ref{eq:aporlar}) into Eq.~(\ref{eq:BH}) yields
\begin{align}
\hat{H}=&\sum_{i}\frac{U}{2}{\delta\hat{n}}_{i}^2-2Jn\cos(\hat{\varphi}_{i}-\hat{\varphi}_{i+1}) +\frac{U}{2}(2n-1){\delta\hat{n}}_{i} \nonumber\\
&+M\frac{U}{2}(n^2-n).
\label{eq:PRfromBH}
\end{align}
Here we neglect the operator $\delta\hat{n}_{i}$ in the hopping terms, because $\sqrt{n+\delta\hat{n}_{i}} \simeq \sqrt{n}+{\delta\hat{n}_{i}}/{(2\sqrt{n})}\simeq \sqrt{n}$. The operator $\sum_{i}\delta\hat{n}_{i}$ still commutes with the Hamiltonian (\ref{eq:PRfromBH}). The substitution $\delta\hat{n}_{i}=\hat{q}_{i}$, $U=E_{\mathrm{C}}$ and $2Jn=E_{\mathrm{J}}$ yields the Josephson junction array Hamiltonian (\ref{eq:PRH}) up to additive renormalizations of the energy and chemical potential, which do not influence the physical properties of the model. Note that we can relax the constraint $\sum_i\hat{q}_i=Q=0$ and consider an arbitrary value of the charge $Q$, which is a conserved quantity.

Thus the model has two integrals of motion but lacks a line of infinite temperatures in the energy-charge plane. If we suppose that the infinite temperature line exists and try to construct the density matrix $\hat{\rho}_0\sim\exp\big(-\alpha \sum_i\hat{q}_i\big)$ by analogy with Sec.~\ref{sec:pert}, we will obtain a divergence in the partition function $Z_0=\tr\exp\big(-\alpha \sum_i\hat{q}_i\big)$, because $q_i$ takes \emph{all integer values} in the Fock basis. The derivation of the model (\ref{eq:PRfromBH})
%, equivalent to (\ref{eq:PRH}),
from the Bose-Hubbard model reveals the reason why: the operators of local charge $\hat{q}_{i} = \delta\hat{n}_i$ are supposed to be \emph{unbounded from below}, while they should actually obey the inequality $-n\leqslant\delta\hat{n}_i$.

\section{Conclusions}

In this work we studied the equilibrium properties of the repulsive quantum Bose-Hubbard model in arbitrary lattice dimensions, with and without disorder.
For the classical limit of a Gross-Pitaevskii lattice in one dimension, the existence of a non-Gibbs phase was proven for two-body interactions \cite{rasmussen00}, and for a case with saturable nonlinearity \cite{samuelsen13}.
We extend these results to the full many-body quantum domain including its classical limit, and to arbitrary lattice dimensions, and to a generalized set of lattice couplings and interaction functions.
We proved the existence of non-Gibbs states in the particle number and energy density control parameter space. The separation line in the density control parameter space between Gibbs and non-Gibbs states $\varepsilon=Un^2$ corresponds to the infinite temperature line where $\beta=0$. We substantially extend these results both for the quantum and classical cases [see Eqs.~(\ref{eq:bGnG}) and (\ref{eq:bGnGcl}) respectively]
for a much wider class of interactions obeying asymptotic functional dependence on the particle density  (\ref{eq:ancond}). This dependence tells us that the upper bound of the Hamiltonian spectrum at given particle density grows faster that the system size in the thermodynamic limit. For this reason, the non-Gibbs phase cannot be cured into a Gibbs one within the standard Gibbs formalism using negative temperatures (see Sec.~\ref{sec:extentions}).

The existence of a non-Gibbs phase needs an infinite temperature line at finite densities in the density control parameter space in the first place.
A prerequisite of such a line is the existence of a density control parameter space, which is at least two-dimensional. Then there needs to be at least one more conserved quantity in addition
to the energy (since we consider Hamiltonian systems). The particle number in the Bose-Hubbard system  is precisely the second integral of motion we need.
Once the infinite temperature line is obtained, a second condition to be satisfied in order for a non-Gibbs phase to exist is the nonsaturability of the interaction potential (\ref{eq:ancond}). This property will
apparently lead to a modification of finite-size scalings of energy and entropy densities. How precisely, remains to be addressed in future work.

From a practical perspective, the Bose-Hubbard lattice is a projection of an infinite band system with unbounded spectrum onto
one (or a few) band(s) in the tight-binding approximation.
The obtained results can be applied to a realistic system with a periodic potential with
some caveats. In the case of one band, the limit of high temperatures actually means that $k_{\mathrm{B}}T\gg J,U$
but $k_{\mathrm{B}}T\ll\Delta$, where $\Delta$ is the gap between
the lowest bands and $k_{\mathrm{B}}$ is the Boltzmann constant.
On the other hand, the chemical potential should be restricted as
well: $|\mu|\ll\Delta$. Then we arrive at the reliable temperature
range
\begin{equation}
\alpha/\Delta\ll\beta\ll\alpha/J,\ \alpha/U,\label{eq:limapp}
\end{equation}
 where the parameter $\alpha$ is directly related to the lattice
filling factor: $\alpha\simeq\ln\left(1+1/n\right)$.

A natural question is about the impact of non-Gibbs states on the dynamics of the system.
For the classical GP lattice case, the non-Gibbs states correlate with an entering of the system into a dynamical glass phase, in which
relaxations of local fluctuations of the particle density are slowing down in a dramatic way \cite{PhysRevLett.120.184101}. This happens
despite the fact that the system shows nicely ergodic (chaotic) spots. It turns out that these spots are however strongly diluted, and the final time to reach ergodicity
is much larger than a typical inverse Lyapunov exponent \cite{1994JSP....76..627K}.
In Ref. \cite{Pino:2016} a similar scenario for a Josephson junction network is expected to lead to the celebrated many-body localization
in the quantum case, which is a kind of Anderson localization in the Fock space.
We therefore expect that the Bose-Hubbard model could
also show a transition into a many-body localized regime, particularly, in the non-Gibbs phase.

\section{Acknowledgement}

This work was supported by the Institute for Basic Science, Project Code
IBS-R024-D1. T.E. acknowledges financial support by the Alexander-von-Humboldt foundation through the Feodor-Lynen Research Fellowship Program
No. NZL-1007394-FLF-P. A.Yu.Ch. thanks the hospitality of the IBS Center for Theoretical Physics of Complex Systems, where most of the work was conducted,  and acknowledges support from the JINR--IFIN-HH projects.

\bibliography{BH2}

\begin{thebibliography}{25}
\bibitem [{not()}]{note}%
  \BibitemOpen
  \href@noop {} {}\bibinfo {note} {In the literature about the discrete
  nonlinear {S}chr\"odinger equation, the notations $\nu=U$, $h=\varepsilon$,
  $A=N$, and $a=n$ are usually used.}\BibitemShut {Stop}%
\end{thebibliography}

\end{document}